\documentclass[12pt]{article}
\usepackage{epsfig}         %standard package. Note graphics<graphicx<epsfig 
\usepackage{url} % allows use of tilde in urls

\newcommand{\beq}{\begin{equation}}  
\newcommand{\eeq}{\end{equation}}
\newcommand{\beqa}{\begin{eqnarray}}  
\newcommand{\eeqa}{\end{eqnarray}}
\newcommand{\rarrow}[0]{\rightarrow}
\newcommand{\Rarrow}[0]{\Rightarrow}

\newcommand{\ul}[1]{\underline{#1}}
\begin{document}

\title{Binary Decision Diagrams are\\
 a Subset of  Bayesian Nets}

\author{Robert R. Tucci\\
        P.O. Box 226\\ 
        Bedford,  MA   01730\\
        tucci@ar-tiste.com}

\date{ \today} 

\maketitle

\vskip2cm
\section*{Abstract}
The goal of this brief pedagogical article is to show that 
Binary Decision Diagrams are a special  kind 
of Bayesian Net. This observation is 
obvious to workers in these two fields, 
but it might not be too obvious to others.

\newpage

The goal of this brief pedagogical article is to show that 
Binary Decision Diagrams\cite{bdd-86}\cite{bdd-intro} are a special  kind 
of Bayesian Net\cite{bnet}. This observation is 
obvious to workers in these two fields, 
but it might not be too obvious to others.

Let $Bool =\{0,1\}$ where $0=false$ and $1=true$. 
For any integer $n$, functions $f:Bool^n\rarrow Bool$ 
are called {\it Boolean expressions}.
Suppose $x,y \in Bool$.
As is conventional, we will use
$x\vee y$ (resp., $x\wedge y$) to denote
the Boolean function of $x$ and $y$ that evaluates to true if 
and only if $x$ OR $y$ (resp., $x$ AND $y$) is true. 
$\overline{x}$ will denote
the opposite of $x$. 

Random variables will be indicated  by underlining. $P(\ul{a} = a)$
will denote the probability that the random variable $\ul{a}$
assumes the value $a$. We will abbreviate 
$P(\ul{a} = a)$ by $P(a)$ if this doesn't lead to confusion.

Let $\delta ^x_y$ denote the Kronecker delta function; it equals 1 if $x=y$, and it 
equals 0 if $x\neq y$.

The {\it if-then-else function} $a(h, c_0, c_1)$ is defined by:

\beq
a(h, c_0, c_1)=
[h\Rarrow (c_0, c_1)]=
[(\overline{h} \wedge c_0)\vee( h\wedge  c_1)]
= \left\{ 
\begin{array}{l}
c_0\;\;{\rm if} \; h=0\\
c_1\;\;{\rm if} \; h=1
\end{array}
\right.
\;.
\eeq
One can interpret $h$ as the hypothesis, $c_0, c_1$ as the two possible
conclusions, and $a$ as the assertion, which may or may not be true,
that if $h$, then $c_1$, else $c_0$.

\begin{figure}[h]
    \begin{center}
    \epsfig{file=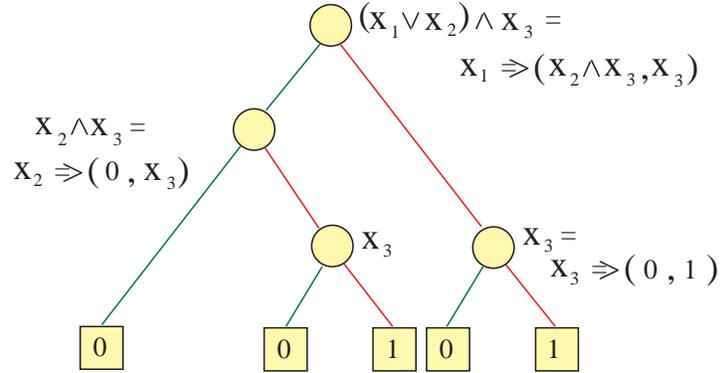, height=2.0in}
    \caption{A decision tree for 
	$(x_1\vee x_2)\wedge x_3$.}
    \label{fig:tree}
    \end{center}
\end{figure}

Note that for any function $f:Bool\rarrow Bool$, one has

\beq
f(x) = 
[x \Rarrow(f(0), f(1))]
\;.
\eeq
This formula is known as the {\it Shannon expansion formula} of $f$ with respect to $x$.
Given any Boolean expression $f(x_1, x_2, \ldots, x_n)$ in $n$ variables,
one can Shannon expand $f$ with respect to $x_1$. This
yields an if-then-else expression whose two conclusions
depend only on the $n-1$ variables  $x_2, x_3, \ldots x_n$.
Next, one can Shannon expand these 2 conclusions
with respect to $x_2$, yielding 
4 conclusion that depend only on $n-2$ variables. 
And so on, until all final conclusions
depend on zero variables, so they are either 0 or 1.
The conclusions from all stages of this process
 can be arranged as a {\it decision tree}.
For instance, consider the following Boolean
expression:

\beq
f(x_1, x_2, x_3) = [(x_1\vee x_2)\wedge x_3)]
\;.
\eeq
Since 

\beqa
[(x_1\vee x_2)\wedge x_3] &=& 
[x_1 \Rarrow (x_2\wedge x_3, x_3)]\;,\\
x_2\wedge x_3 &=& [x_2 \Rarrow(0, x_3)]\;,\\
x_3 &=& [x_3 \Rarrow (0,1)]
\;,
\eeqa
we get the decision tree of Fig.\ref{fig:tree}.
Some nodes will occur more than once. By merging 
each set of equivalent  nodes into a single node, we can transform
the decision tree into a directed acyclic graph (dag).
These dags, arising from a 
recursive application of the Shannon expansion formula, 
 were invented  by R. Bryant in Ref.\cite{bdd-86}. He called them
BDDs (or, more precisely, reduced ordered BDDs).

\begin{figure}[h]
    \begin{center}
    \epsfig{file=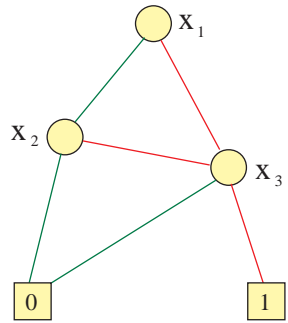, height=1.7in}
    \caption{A BDD for $(x_1\vee x_2)\wedge x_3$.}
    \label{fig:bdd}
    \end{center}
\end{figure}

\begin{figure}[h]
    \begin{center}
    \epsfig{file=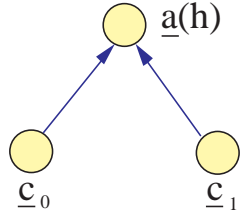, height=1.2in}
    \caption{ If-then-else node.}
    \label{fig:if-then-else}
    \end{center}
\end{figure}

\begin{figure}[h]
    \begin{center}
    \epsfig{file=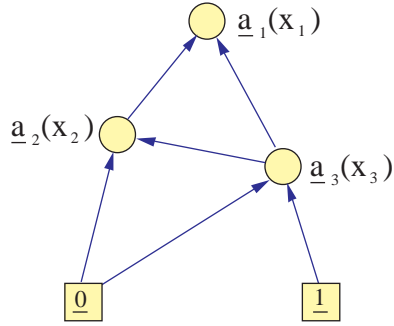, height=1.7in}
    \caption{A Bayesian net for 
$(x_1\vee x_2)\wedge x_3$.}
    \label{fig:bnet}
    \end{center}
\end{figure}

In the field of Bayesian nets, one can define an if-then-else node.
See Fig.\ref{fig:if-then-else}. The node is labelled 
by a random variable $\ul{a}(h)$ which depends on a Boolean parameter $h$.
The node has two possible states, $0$ and $1$.
The transition probability associated with node $\ul{a}(h)$
is given by 

\beq
P\{\underline{a}(h) = a | c_0, c_1\} = \delta^a_{h\Rarrow (c_0, c_1)}
\;,
\eeq
for all $h, a, c_0, c_1\in Bool$.
This transition probability can be expressed in table form 
as follows. For $h=0$, one has

\beq
P\{\underline{a}(0) = a | c_0, c_1\} =
\left\{
\begin{array}{cc}
& (c_0, c_1)\rarrow\\
\begin{array}{c}a\\ \downarrow\end{array}&
\begin{array}{c|cccc}
 & 00 &01 &10 &11\\
\hline
0&1&1&0&0\\
1&0&0&1&1
\end{array}
\\
\end{array}
\right.
\;,
\eeq
and for $h=1$,

\beq
P\{\underline{a}(1) = a | c_0, c_1\} =
\left\{
\begin{array}{cc}
& (c_0, c_1)\rarrow\\
\begin{array}{c}a\\ \downarrow\end{array}&
\begin{array}{c|cccc}
 & 00 &01 &10 &11\\
\hline
0&1&0&1&0\\
1&0&1&0&1
\end{array}
\\
\end{array}
\right.
\;.
\eeq
Note that these probabilities are all {\it deterministic},
meaning that they are all either 0 or 1, none is a fraction.

The BDD of Fig.\ref{fig:bdd} 
is equivalent to the Bayesian Net of Fig.\ref{fig:bnet}.
Note that these two dags are identical.
In the Bayesian net of Fig.\ref{fig:bnet}, all nodes are if-then-else nodes except for 
the square ones. Square nodes have two states, 0 and 1. Node $\ul{0}$
(resp., node $\ul{1}$)
is in state 0 (resp., state 1) with probability 1, and in the opposite state with probability 0.

It is now clear that BDDs are just a special kind of Bayesian Net.

{\footnotesize
\begin{tabular}{|l|l|l|}
\hline 
				&{\bf BDD} & {\bf Bayesian Net}\\
\hline 
{\bf number of states} 	& 2		& arbitrary		\\
{\bf of a node}		& 			&			\\
\hline 
{\bf number of arrows} 	& 2 if round node,			& arbitrary		\\
{\bf entering a node}	& 0 if square node				&			\\
\hline
{\bf transition probabilities} &deterministic		& arbitrary		\\	
{\bf for a node}		& 				&			\\
\hline
{\bf number of  nodes} 	& 1				& arbitrary		\\
{\bf at top  of graph }	&				&			\\	
\hline
{\bf number of  nodes} 	& 2				& arbitrary		\\
{\bf at bottom  of graph }	&				&			\\	
\hline
\end{tabular}
}


\begin{thebibliography}{99}
\bibitem{bdd-86}The field of BDDs started with the following paper:
``Graph Based Algorithms for Boolean Function Manipulation",
by Randal E. Bryant, IEEE Trans. on Computers, 8(C-35)677, 1986.
This paper is available at Bryant's website:\newline
\url{http://www-2.cs.cmu.edu/~bryant/}

\bibitem{bdd-intro}
``An Introduction to Binary Decision Diagrams", by Henrik Reif Andersen.
Available at Andersen's website:\newline
\url{http://www.itu.dk/people/hra/}

\bibitem{bnet}
For an introduction to Bayesian Nets, see, for example,
this web site:\newline
\url{http://www.cs.berkeley.edu/~murphyk/Bayes/bayes.html}


\end{thebibliography}
\end{document}